# Nanoscale Communication with Brownian Motion

Andrew W. Eckford
Department of Computer Science and Engineering, York University
4700 Keele Street, Toronto, Ontario, Canada M3J 1P3
E-mail: aeckford@yorku.ca

*Abstract*— **In this paper, the problem of communicating using chemical messages propagating using Brownian motion, rather than electromagnetic messages propagating as waves in free space or along a wire, is considered. This problem is motivated by nanotechnological and biotechnological applications, where the energy cost of electromagnetic communication might be prohibitive. Models are given for communication using particles that propagate with Brownian motion, and achievable capacity results are given. Under conservative assumptions, it is shown that rates exceeding one bit per particle are achievable.**

## I. Introduction

In most existing forms of engineered communication, messages are transmitted over electromagnetic carriers. Although this form of communication has been remarkably successful, the emerging field of *nanotechnology* poses communication challenges for which electromagnetic communication might be unsuitable. For example, in a conducting fluid (such as blood, or seawater), electromagnetic waves cannot propagate, while alternatives (such as sonar) may be problematic for very small devices. Furthermore, electromagnetic communication generally imposes an energy cost which might be undesirable.

In nature, it is very well understood that *chemical communication* is used for communication between nanoscale "machines", such as cells or microbes. This form of communication is desirable in biological systems owing to its simplicity and low energy cost. One such method is *quorum sensing*, in which bacteria exchange messages intended to determine roughly the local population of their species [1]. This form of communication has attracted the attention of engineers: in [2], the genetic sequences of these communication components were isolated, with the intention of using them to allow communication and co-operation between engineered microbial "robots"; or to force them to carry out chemical functions analogous to logic gates [3]. Recent work has attempted to characterize this pathway as a linear communication channel [4].

Generally, the biological literature has attempted to explain the function of chemical messaging, rather than exploiting it for artificial purposes. Our contribution in this paper is to obtain models for chemical communication channels, and give achievable capacity values for those channels. As such, the purpose of this paper is to determine the feasibility of this type of communication in nanoscale devices. These channels are essentially timing channels, in which the noise is the delay between releasing a particle into the medium and observing its arrival, so previous work on queue timing channels [5], [6] is closely related. Furthermore, work on diffusion channels has been carried out by Berger (e.g., [7]), though the aim of his work is to analyze biochemical processes through the lens of information theory.

The chemical channel is a practically interesting system which is poorly understood from the perspective of communication. In particular, it is currently unknown how to model this channel, and it is therefore useful to know its physical limits in terms of information-theoretic capacity. Furthermore, even though the computational capabilities of very tiny machines are currently rudimentary (which restricts the use of coding, or complicated modulation), the capacity gives a very loose upper bound on the uncoded capabilities of the system, and gives a rough idea as to the potential of chemical communication.

## II. Model

### A. Basic assumptions

We consider a chemical communication system as in Figure 1. The transmitter has a reservoir of particles, and forms messages by releasing particles at a vector of transmission times $\mathbf{x} = [x_1, x_2, \ldots, x_\ell]$, where $x_i$ is the time of release of the $i$th particle, and $\ell$ is the total number of particles released to convey the message. There is a distance $d > 0$ between the transmitter and the receiver. On release, each particle enters a fluid medium between the transmitter and receiver, and the position of the $i$th particle at any time $t > x_i$ is given by a Brownian motion $B_i(t)$.

The following are the key assumptions of this system:

- The transmitter perfectly controls the departure time of each particle. After release, the transmitter is "transparent" to the particles, so that there is no effect on the particles if they cross the origin.
- The particle propagates until the *first hitting time* at the receiver (i.e., the smallest $t$ such that $B_i(t) = d$). The receiver perfectly observes the hitting time and removes the particle from the system.
- The medium between the transmitter and receiver is one-dimensional, with the transmitter at the origin and the receiver at $d > 0$. The medium is semi-infinite, defined on $(-\infty, d]$. (Particles can never achieve a position greater than $d$, since they are removed at their first hitting time.)
- For particles $i$ and $j$, the paths $B_i(t)$ and $B_j(t)$ are independent if $i \neq j$.

This is an idealized channel model which simplifies the system and eliminates all possible sources of noise other than the transmission time.

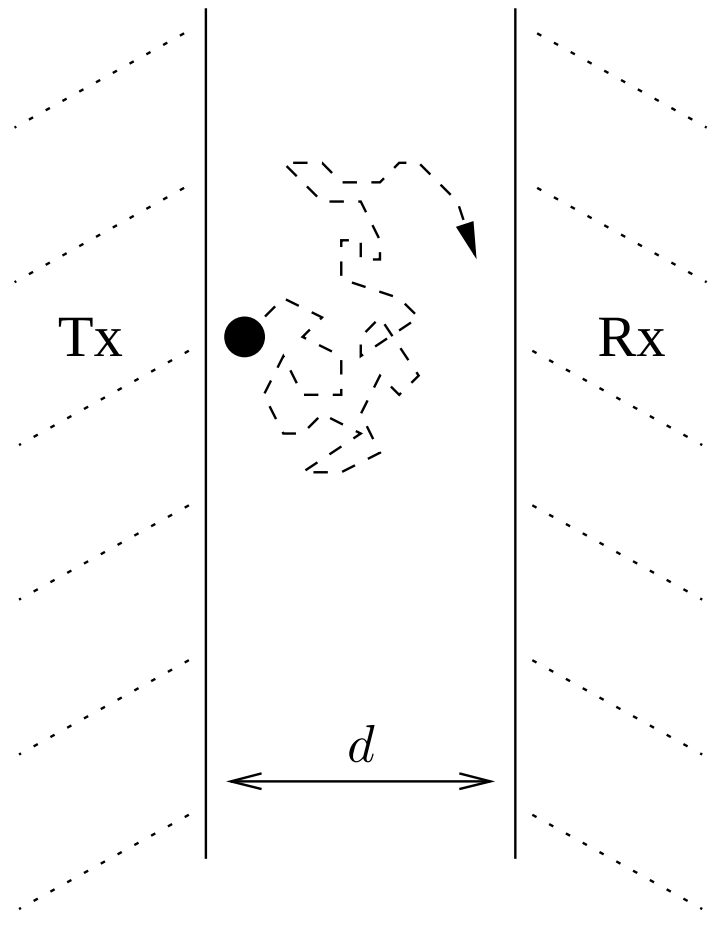

Fig. 1. A chemical communication system. A particle is propagating from transmitter (Tx) to receiver (Rx) using Brownian motion.

The *transmission time* $t_i$ of each particle is a random variable defined as the length of time between release from the transmitter and the first hitting time at the receiver. If the Brownian motion channel is considered to be a timing channel, then $f_{T_i}(t_i)$ can be thought of as the PDF of the noise process. The distribution of the first hitting time is the *only stochastic property of Brownian motion* that we require for this analysis.

### B. Channel input-output relationship

From Section II-A, $\mathbf{x}$ represents a vector of particle release times, and the particle released at time $x_i$ arrives at time $x_i+t_i$. Letting $\mathbf{t} = [t_1, t_2, \ldots, t_n]$ represent the vector of transmission times for each particle, we can form the vector

$$\mathbf{u} := \mathbf{x} + \mathbf{t}, \tag{1}$$

where $u_i = x_i + t_i$ represents the first hitting time of the $i$th particle at the detector.

However, the detector does not observe $\mathbf{u}$ directly, because $\mathbf{u}$ is stated in the order that the particles were released, which is not necessarily the same as the order in which the particles hit the detector. Instead, the detector observes

$$\mathbf{y} := \mathrm{sort}(\mathbf{u}), \tag{2}$$

where the function $\mathrm{sort}(\cdot)$ takes a vector argument, and returns the vector sorted in increasing order.

Suppose the particles are distinguishable – that is, every particle carries a unique label. (In this paper, we assume that the transmitter always releases labeled particles in the same order, so the labels carry no information related to $\mathbf{x}$, but this assumption can be relaxed.) Then the detector observes the pair of vectors $(\mathbf{y}, \mathbf{b})$, where $\mathbf{b}$ is a vector of labels, and where, for all $i$, $b_i$ is the label attached to the particle that arrived at time $y_i$.

Obviously, if every element of $\mathbf{b}$ is unique, the detector can use the pair $(\mathbf{y}, \mathbf{b})$ to recover the vector $\mathbf{u}$ from (1). In this case, since the transmission times $t_i$ are all independent, the channel is equivalent to an additive noise channel. We write $\pi(\mathbf{y}, \mathbf{b})$ as the inverse of the sorting operation, so that $\mathbf{u} = \pi(\mathbf{y}, \mathbf{b})$ is the permutation of $\mathbf{y}$ that restores the original order of the labels. Thus, we have that

$$f(\mathbf{y}, \mathbf{b} \mid \mathbf{x}) = \prod_{i=1}^{n} f_{T_i}(u_i - x_i), \tag{3}$$

so long as $\mathbf{y}$ is in increasing order (the probability is zero otherwise). This channel may be handled in the same manner as any additive noise channel, and we give some example capacity calculations in Section IV.

Now suppose the particles are *indistinguishable*, so that $\mathbf{b}$ is not available to the detector. In the following example, we derive the PDF $f(\mathbf{y} \mid \mathbf{x})$, from first principles, for two indistinguishable particles:

*Example 1:* Let $\mathbf{x} = [x_1, x_2]$ represent the release times of two indistinguishable particles, and let $\mathbf{y} = [y_1, y_2]$ represent the first hitting times of these particles, sorted in order of arrival at the detector, as in (2). Then we may write

$$\begin{aligned} & f(\mathbf{y} \mid \mathbf{x}) \\ & \quad = \quad f(y_1, y_2 \mid x_1, x_2) \\ & \quad = \quad f(y_1, y_2 | x_1, x_2, u_1 < u_2) \Pr(u_1 < u_2 | x_1, x_2) + \\ & \qquad\quad f(y_1, y_2 | x_1, x_2, u_1 \geq u_2) \Pr(u_1 \geq u_2 | x_1, x_2). \end{aligned} \tag{4}$$

Now consider $f(y_1, y_2|x_1, x_2, u_1 < u_2)$. If we know that $u_1 < u_2$, then we know that $\mathbf{y} = \mathrm{sort}([u_1, u_2]) = [u_1, u_2]$, so $u_1 = y_1$ and $u_2 = y_2$. Thus,

$$\begin{aligned} & f(y_1, y_2 | x_1, x_2, u_1 < u_2) = \\ & \qquad f_{U_1, U_2 | X_1, X_2, U_1 < U_2}(y_1, y_2 | x_1, x_2, u_1 < u_2). \end{aligned} \tag{5}$$

Furthermore, from Bayes' rule,

$$\begin{aligned} & f(u_1, u_2 | x_1, x_2, u_1 < u_2) \\ & \quad = \begin{cases} \frac{f(u_1, u_2 | x_1, x_2)}{\Pr(u_1 < u_2 | x_1, x_2)}, & u_1 < u_2; \\ 0, & u_1 \geq u_2. \end{cases} \end{aligned} \tag{6}$$

Thus, from (5) and (6),

$$\begin{aligned} & f(y_1, y_2 | x_1, x_2, u_1 < u_2) = \\ & \qquad \begin{cases} \frac{f_{U_1, U_2 | X_1, X_2}(y_1, y_2 | x_1, x_2)}{\Pr(u_1 < u_2 | x_1, x_2)}, & y_1 < y_2; \\ 0, & y_1 \geq y_2. \end{cases} \end{aligned} \tag{7}$$

By a similar argument, we can write

$$\begin{aligned} & f(y_1, y_2 | x_1, x_2, u_1 \geq u_2) = \\ & \qquad \begin{cases} \frac{f_{U_1, U_2 | X_1, X_2}(y_2, y_1 | x_1, x_2)}{\Pr(u_1 \geq u_2 | x_1, x_2)}, & y_2 \geq u_1; \\ 0, & y_2 < y_1. \end{cases} \end{aligned} \tag{8}$$

Substituting (7) and (8) into (4), we can write

$$\begin{aligned} & f(\mathbf{y} \mid \mathbf{x}) = \\ & \qquad \begin{cases} f_{U_1, U_2 | X_1, X_2}(y_1, y_2 | x_1, x_2) + \\ \; f_{U_1, U_2 | X_1, X_2}(y_2, y_1 | x_1, x_2), & y_1 \leq y_2; \\ \\ \qquad\qquad 0, & y_1 > y_2. \end{cases} \end{aligned} \tag{9}$$

*(End of example.)*

Returning to (3), we see that the same expression is found by taking the sum over all possible values of $\mathbf{b}$:

$$\begin{aligned} f(\mathbf{y} \mid \mathbf{x}) &= \sum_{\mathbf{b} \in \mathcal{P}} f(\mathbf{y}, \mathbf{b} \mid \mathbf{x}) \\ &= \begin{cases} \sum_{\mathbf{b} \in \mathcal{P}} f_{\mathbf{U}}(\pi(\mathbf{y}, \mathbf{b}) \mid \mathbf{x}), & \mathbf{y} = \mathrm{sort}(\mathbf{y}), \\ 0, & \mathbf{y} \neq \mathrm{sort}(\mathbf{y}); \end{cases} \end{aligned} \tag{10}$$

where $\mathcal{P}$ represents all possible permutations of $n$ letters. In the case where $n = 2$, as in Example 1, there are only two possible permutations, and we immediately see that (10) is equivalent to (9).

It can be shown that exact calculation of the PDF in (10) is equivalent to taking the *permanent* of an $n \times n$ matrix. Calculating the permanent is known to be a member of the class of #P-complete problems[1] [8], which are known to be intractable for large $n$.

### C. Discrete-time model

Instead of observing the exact arrival times of each particle, suppose we have a *discrete-time* model with the following properties:

- Time is partitioned into intervals, indexed by $\mathcal{I} = \{1, 2, \ldots, \ldots, |\mathcal{I}|\}$, each of duration $\tau$.
- Particles are only released at the beginning of an interval. For $i \in \mathcal{I}$, the vector $\mathbf{r} = [r_1, r_2, \ldots, r_{|\mathcal{I}|}]$ gives the number of particles released at the beginning of each interval, where for $i \in \mathcal{I}$, $r_i$ represents the number of particles released at the beginning of the $i$th interval.
- The detector reports the *count* of the number of particles that arrive on each interval. The vector $\mathbf{c} = [c_1, c_2, \ldots, c_{|\mathcal{I}|}]$ gives the counts in each interval, where for $i \in \mathcal{I}$, $c_i$ represents the number of particles that arrived in the $i$th interval.

This model is no less intractable as compared to the continuous-time model. However, we will see in Section III that a reasonably good (and tractable) approximation exists for this model, which leads to a lower bound on the capacity of the system. Even for the exact discrete-time model, it is obvious that such a model leads to a lower bound on the system capacity.

### D. Statistical model of transmission time

To model the diffusion process from the transmitter to the receiver, we use the *Wiener process*. There are better physical models for Brownian motion, but the Wiener process has the advantage that the PDF of the first hitting time $t$ of each particle can be expressed in closed form. In the remainder of the paper, none of the techniques depend on this particular PDF for the first hitting time $t$, so it changes nothing to substitute it for any other model for the first hitting time, or to include such things as a Brownian motion with drift.

A Wiener process $w(t)$ is a continuous-time random process where, for $t' > t$ and for some constant $\sigma^2$, $w(t') - w(t)$ is Gaussian distributed with zero mean and variance $\sigma^2(t' - t)$; and where the increment $w(t') - w(t)$ on the interval $[t, t']$ is independent of the increment on any other disjoint interval. We assume that $w(0) = 0$, and that $w(t)$ is undefined for $t < 0$. It is a well-known result for the Wiener process (see, e.g., [9]) that this first hitting time (i.e., the transmission time), written $t_i$ for the $i$th particle, has a PDF given by

$$f(t_i) = \begin{cases} 0, & t_i \leq 0, \\ \frac{d}{\sqrt{2\pi\sigma^2 t_i^3}} \exp\left(-\frac{d^2}{2\sigma^2 t_i^2}\right), & t > 0. \end{cases} \tag{11}$$

From (11), $f(t_i)$ has an extremely long tail that decays as $\Theta(t_i^{-3/2})$. The mean, and all other moments of this density, are equal to $\infty$. As a result, if a detector waits for all particles to arrive before decoding a message, the average waiting time will be $\infty$, which means that the average data rate, in bits per second, could be zero. In such a case, it may make sense to define a transmission interval $T$, and declare any particle with transmission time $t_i > T$ to be lost.

The constants $d$ and $\sigma^2$ depend on the physical properties of the system. In the remainder of the paper, we will assume for simplicity that $d = \sigma^2 = 1$.

## III. Capacity bounds

### A. Simplified systems

We firstly consider the following simplified systems, calculating capacity in bits per unit time for an unbounded number of particles, and capacity in bits per particle for unbounded time. In both cases the capacity is infinite:

- **Unbounded number of particles.** The particle-release channel is an infinite-server queue, so we can use a similar argument to the calculation of the infinite-server queue capacity [6] to show that its capacity, in bits per unit time, is $\infty$.
- **Unbounded time.** We can take an interval of time $T$ and divide it into segments of length $\log T$. Using pulse-position modulation, a message is sent by transmitting a *single* particle at the beginning of one of the $T/\log T$ segments. As $T \to \infty$, the particle arrives within the same segment with $\Pr = 1$, allowing the error-free transmission of $\log_2(T/\log T)$ bits; and since $T/\log T \to \infty$ as $T \to \infty$, so does $\log_2(T/\log T)$, so the capacity, in bits per particle, is $\infty$.

Since both time *and* particles are precious resources, one might consider capacity per unit time and per particle. Furthermore, releasing an enormous number of particles at once is impractical, so we can consider limitations on the transmission rate of the particles (e.g., the transmitter is allowed to release at most one particle per unit time). We will consider both of these circumstances in Section IV.

### B. Labeled particles

As we indicated in Section II, the calculation of $f(\mathbf{y} \mid \mathbf{x})$ is intractable. However, if the vector $\mathbf{b}$ of permuted labels is observed, then $f(\mathbf{y}, \mathbf{b} | \mathbf{x})$ is both tractable and straightforward. From (3), knowledge of $\mathbf{b}$ and $\mathbf{y}$ recovers $\mathbf{u}$, and separates

[1] #P-complete is pronounced "sharp-P complete".

each particle into an independent channel with input $x_i$ and output $u_i$. Thus, $I(\mathbf{Y},\mathbf{B};\mathbf{X})$ can be calculated straightforwardly, as for any additive independent channel.

The operation of "labeling" a particle might be costly. For instance, it might be accomplished by maintaining a reservoir of unique particles, or by synthesizing a novel particle for each element of $\mathbf{x}$. As a result, we can consider labellings that use fewer unique elements. For instance, suppose every *second* particle has a unique label. Now, the vector $\mathbf{b}$ does not exactly recover $\mathbf{u}$, but partitions the vector $\mathbf{y}$ into independent channels containing *pairs* of indistinguishable particles, but where the pair of particles in each channel is distinguishable from the particles in every other channel. Such a scheme would use half as many labels as a scheme where every particle is uniquely labeled.

We use the notation $\mathbf{b}^{(j)}$ to indicate that every $j$th label in the vector is unique. That is, as $n \to \infty$, $\mathbf{b}^{(j)}$ contains $n/j$ unique labels. To be consistent with our notation from Section II, we let $\mathbf{b}^{(1)} := \mathbf{b}$. We will calculate some example capacities for such channels in Section IV, but the following proposition gives a straightforward ordering of labellings in terms of mutual information:

*Proposition 1:* If $j < k$, then $I(\mathbf{Y},\mathbf{B}^{(j)};\mathbf{X}) \geq I(\mathbf{Y},\mathbf{B}^{(k)};\mathbf{X})$.

*Proof:* Suppose the total number of particles is $n$. Consider a labeling $\mathbf{b}^{(j)}$ corresponding to the sequence of first hitting times $\mathbf{y}$, recalling that $\mathbf{b}^{(j)}$ contains $n/j$ unique labels. Suppose, between the transmitter and receiver, there is an entity that modifies the labels (without modifying the particle trajectories), as follows. A fraction $n/j - n/k$ of labels are selected, uniformly at random from all possible such selections, (leaving $n/k$ labels unselected); the particles in these labels are then divided (uniformly at random) into $n/k$ groups, corresponding to the unselected labels. The labels on the particles are then replaced with a label from the $n/k$ unselected labels, such that each group receives a unique unselected label. The result is a labeling $\mathbf{b}^{(k)}$ with $n/k$ unique labels. In the limit as $n \to \infty$, the effect of non-integer quotients from any of these divisions is negligible.

Since the relabeling process is a reversible physical process, which is independent of $\mathbf{x}$, the system with labeling $\mathbf{b}^{(k)}$ is physically degraded with respect to the system with labeling $\mathbf{b}^{(j)}$, which is sufficient to prove the proposition. ■

An obvious corollary of Proposition 1 is that the system in which every particle is distinguishable has the largest capacity of any possible such system. Also, since this is a mutual information result rather than a capacity result, it is true for any possible input distribution.

### C. Bounds from approximate PDFs

As we have seen, calculating the exact PMF of the random process $\mathbf{y}$ is intractable for any practical number of particles. However, the process is straightforward to generate: given a vector $\mathbf{x}$ of release times, we simply generate random transmission times for each particle, and sort the result in increasing order. Thus, performing *monte carlo* expectations of any tractable function of $\mathbf{y}$ can be accomplished with reasonable complexity.

The mutual information between the random variables $\mathbf{X}$ and $\mathbf{Y}$ can be written

$$I(\mathbf{X};\mathbf{Y}) = E\left[\log \frac{f(\mathbf{y},\mathbf{x})}{f(\mathbf{y})f(\mathbf{x})}\right]. \tag{12}$$

Of course, taking the *monte carlo* expectation of this function results in no complexity advantage, since the function is still intractable.

However, suppose we replace $f(\mathbf{y},\mathbf{x})$ with a *tractable approximation* $g(\mathbf{y},\mathbf{x})$, with the following properties:

- $\int_{\mathbf{x}}\int_{\mathbf{y}} g(\mathbf{y},\mathbf{x})d\mathbf{y}d\mathbf{x} = 1$ (i.e., $g(\mathbf{y},\mathbf{x})$ is a PDF); and
- $\int_{\mathbf{y}} g(\mathbf{y},\mathbf{x})d\mathbf{y} = f(\mathbf{x})$ (i.e., the correct marginal distribution of $\mathbf{x}$ is preserved).

Then we could write

$$I(\mathbf{X};\mathbf{Y}) \approx E\left[\log \frac{g(\mathbf{y},\mathbf{x})}{\left(\int_{\mathbf{x}} g(\mathbf{y},\mathbf{x})d\mathbf{x}\right) f(\mathbf{x})}\right], \tag{13}$$

and, since $g(\mathbf{y},\mathbf{x})$ is tractable, it would be possible to calculate this approximation to $I(\mathbf{X};\mathbf{Y})$ using *monte carlo* methods.

In fact, we can show that the approximation in (13) is a *lower bound*:

*Proposition 2:* For any PDF $g(\mathbf{y},\mathbf{x})$ satisfying the above properties,

$$I(\mathbf{X};\mathbf{Y}) \geq E\left[\log \frac{g(\mathbf{y},\mathbf{x})}{\left(\int_{\mathbf{x}} g(\mathbf{y},\mathbf{x})d\mathbf{x}\right) f(\mathbf{x})}\right], \tag{14}$$

with equality if and only if $f(\mathbf{y},\mathbf{x}) = g(\mathbf{y},\mathbf{x})$.

*Proof:* We can rewrite (14) as

$$\begin{aligned} &E\left[\log \frac{g(\mathbf{y},\mathbf{x})}{\left(\int_{\mathbf{x}} g(\mathbf{y},\mathbf{x})d\mathbf{x}\right) f(\mathbf{x})}\right] \\ &= H(\mathbf{X}) + E\left[\log g(\mathbf{x} \mid \mathbf{y})\right] \\ &= H(\mathbf{X}) - H(\mathbf{X} \mid \mathbf{Y}) - D\left(f(\mathbf{x}|\mathbf{y}) \parallel g(\mathbf{x}|\mathbf{y})\right) \\ &= I(\mathbf{X};\mathbf{Y}) - D\left(f(\mathbf{x}|\mathbf{y}) \parallel g(\mathbf{x}|\mathbf{y})\right), \end{aligned} \tag{15}$$

where $D(f \parallel g)$ represents Kullback-Leibler (KL) divergence. The proposition immediately follows from (15) and the properties of KL divergence. ■

Since Proposition 2 gives a lower bound on mutual information, it also gives a lower bound on capacity for any input distribution $f(\mathbf{x})$.

### D. Approximate discrete time model

In Proposition 2, *any* PDF satisfying the given properties can be used. However, (15) tells us to look for an approximate PDF that minimizes the KL divergence to the true density $f(\mathbf{y},\mathbf{x})$ (or, in the case of the discrete time model, $(\mathbf{c},\mathbf{r})$). Thus, it is reasonable to look for a tractable density that reasonably approximates $f(\mathbf{c},\mathbf{r})$.

We can modify the discrete time model from Section II-C as follows. Suppose a single particle is transmitted at the beginning of an interval $\tau$. Its probability of arriving during that interval is given by

$$p_{\mathrm{arr}} = \int_{t=0}^{\tau} f(t)dt,$$

where $f(t)$ is specified in (11). Thus, the probability that the particle will arrive in a different interval is given by $1 - p_{\mathrm{arr}}$.

In the $i$th interval, the discrete-time counting detector forms the observation

$$c_i = \hat{r}_i + z_i, \tag{16}$$

where, assuming at most one particle is released,

$$\Pr(\hat{r}_i = 1) = \begin{cases} 0, & r_i = 0, \\ p_{\mathrm{arr}}, & r_i = 1. \end{cases} \tag{17}$$

and where $z_i$ is a Poisson-distributed random variable with arrival rate

$$\lambda = E[r_i](1 - p_{\mathrm{arr}}).$$

In other words, $z_i$ is a "background" arrival rate for particles in the system, as an average of $E[r_i](1 - p_{\mathrm{arr}})$ particles will arrive as a result of particles that did not arrive in the interval in which they were transmitted. A similar model was used to approximate the process of corn pollen dispersal [10].

We can modify this model in an interesting way to achieve higher fidelity. The probability that a particle will arrive in the $k$th interval after its transmission is given by

$$p_{\mathrm{arr}}^{(k)} = \int_{t=(k-1)\tau}^{k\tau} f(t)dt.$$

Let $\hat{r}_i^{(k)}$ represent the analog of the previously defined $\hat{r}_i$, where

$$\Pr(\hat{r}_i^{(k)} = 1) = \begin{cases} 0, & r_i = 0, \\ p_{\mathrm{arr}}^{(k)}, & r_i = 1. \end{cases}$$

Now, the counting process $c_i$ is given by

$$c_i = z_i + \sum_{j=0}^{N-1} \hat{r}_{i-j}^{(j+1)},$$

and $z_i$ is a Poisson-distributed random variable with arrival rate

$$\lambda = E[r_i] \left( 1 - \sum_{j=1}^{N} p_{\mathrm{arr}}^{(j)} \right)$$

In other words, this channel has a sense of intersymbol interference. These models can be easily generalized to the case where more than one particle is released at the beginning of an interval.

## IV. Examples

In this section we present two examples of achievable results, under conditions suggested in previous sections.

*Example 2:* Consider a system with labeled particles and no rate restrictions on particle release. Our strategy is as follows: release each labeled particle on the interval $[0, T]$, and wait until $T$ for particles to arrive; if they have not yet arrived by time $T$, the particles are declared missing. We calculate $I(\mathbf{Y}, \mathbf{B}; \mathbf{X})/T$ to obtain the capacity, per particle per second, for the case where every particle is labeled uniquely. For comparison (and to demonstrate Proposition 1), we also include a case for $I(\mathbf{Y}, \mathbf{B}^{(2)}; \mathbf{X})/T$. In both cases, we consider uniform transmission of particles on the interval $[0, T]$.

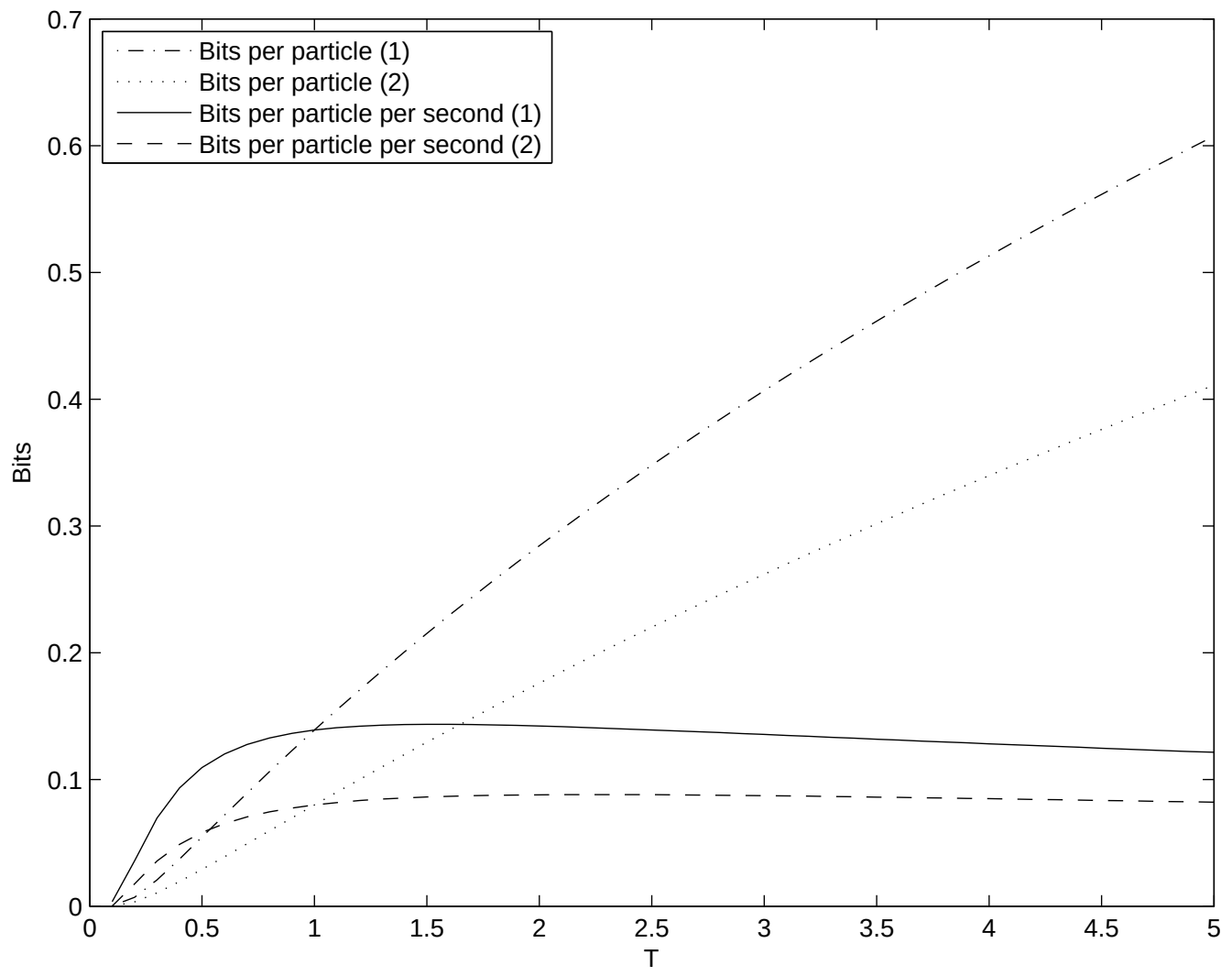

Fig. 2. Mutual information results for rate-unlimited systems. Systems designated (1) have unique labels for every particle, and systems designated (2) have unique labels for every second particle.

Results are depicted in Figure 2. In the figure, we see that the mutual information per particle increases monotonically with $T$, as expected, but that the mutual information per particle per second reaches a maximum value. Furthermore, as expected, the labeling $\mathbf{b}^{(2)}$ has smaller mutual information than the labeling $\mathbf{b}$. *(End of example.)*

In future work, we will optimize the input distribution $p(x)$, but preliminary results indicate that the optimized distribution is probably close to the uniform distribution.

In Example 2, we assumed that particles were distinguishable, and that there was no restriction on the number of particles released in an interval of time. Thus, Example 2 is calculated on the assumption that an infinite number of distinguishable particles (or distinguishable pairs of particles) are released in the time interval $[0, T]$. In the following example, we use Proposition 2 and the approximate discrete-time model from Section III-D to present an achievable mutual information result under a (more practical) constraint on the rate of particle release.

*Example 3:* Suppose our system operates under an average-case particle release constraint that, on average, at most five particles can be released per second. To achieve this constraint using the discrete-time model, we will release at most one particle per interval, and select $\tau$ and $p(r)$ accordingly (we will assume that the probability of transmitting a symbol is independent from interval to interval).

We use the "inter-symbol interference" approximate model with $N = 2$, which appears to have the best results of this class of models. In Figure 3, we present a lower bound on mutual information per second, and in Figure 4, we present a lower bound on mutual information per particle, in both cases using *monte carlo* expectation and Proposition 2. As expected, the capacity per particle is highest when $p(r)$ is small, meaning that there are very few particles in the system (this follows from our argument in Section III-A). However, the capacity per unit time is small when $p(r)$ is small. Also, as expected, the bound on mutual information per unit time increases as $\tau$ increases, but reaches a maximum around $\tau = 1$, representing the balance between discernibility of the particles and the long interval between particles. *(End of example.)*

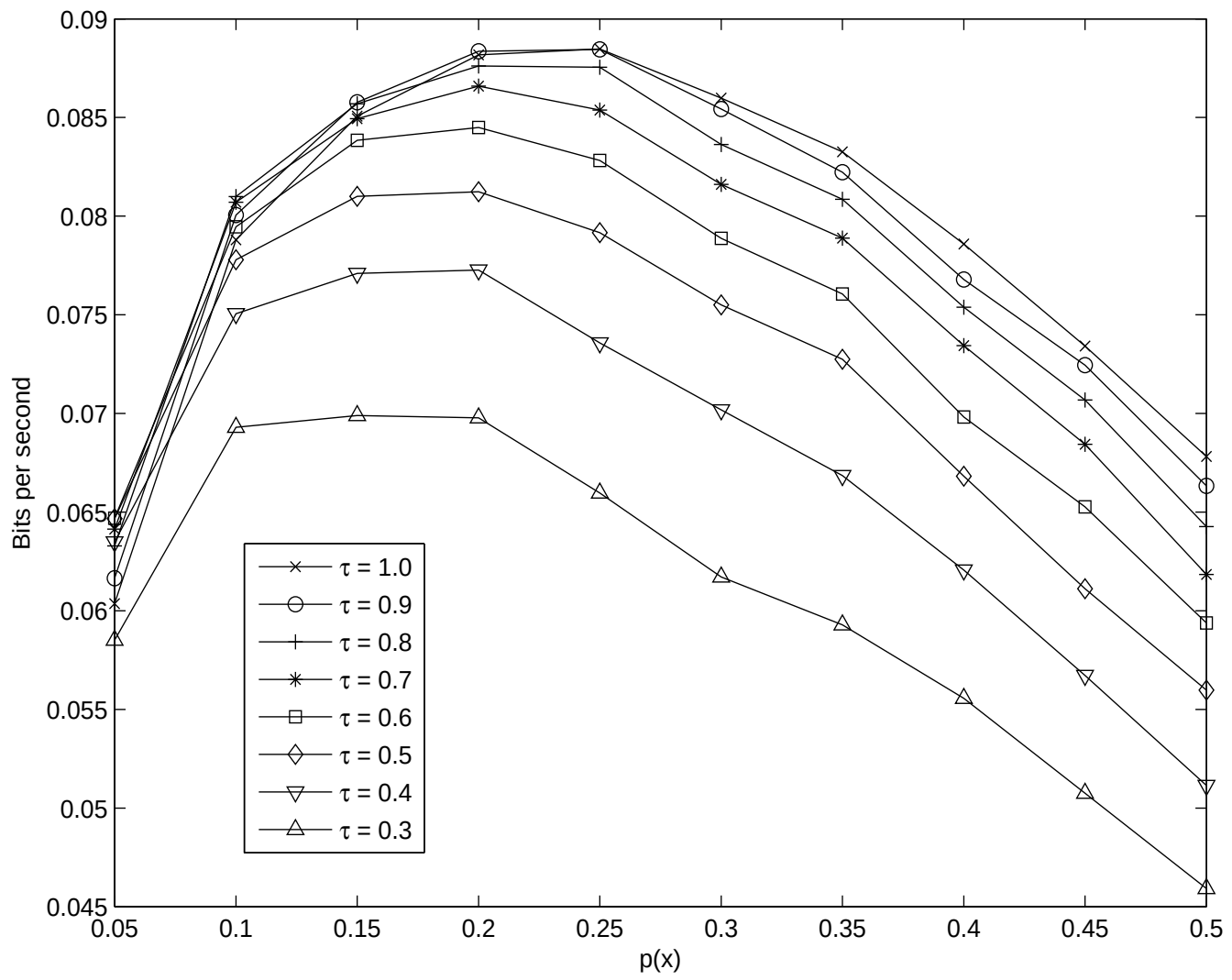

Fig. 3. Mutual information per second with respect to $p(x)$ for various values of $\tau$.

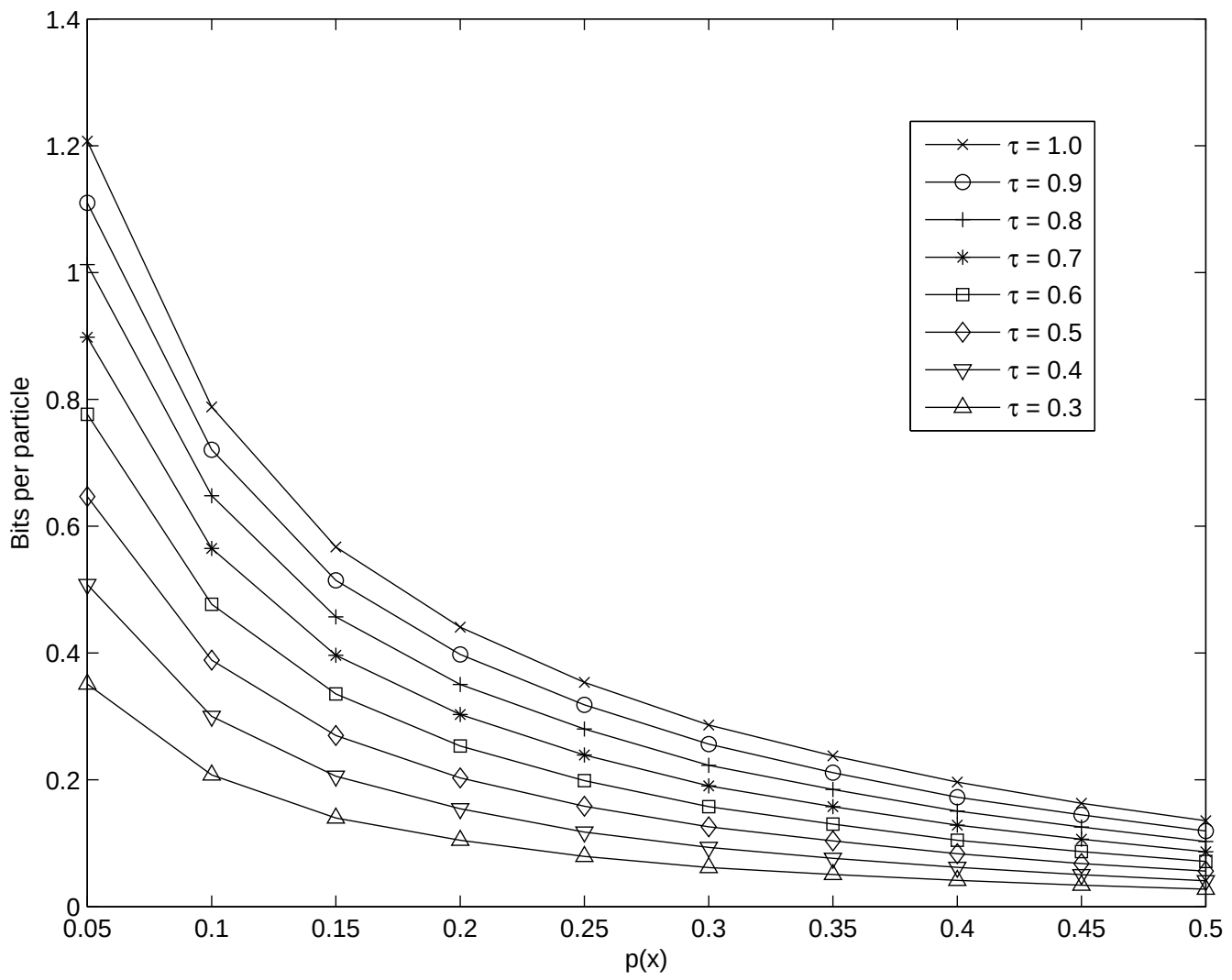

Fig. 4. Mutual information per particle with respect to $p(x)$ for various values of $\tau$.

A remarkable consequence of Example 3 and Figure 4 is that, under practical assumptions and at the maximum rate of bits per second, transmission of a message of $k$ bits requires roughly $3k$ particles. Thus, considering a system where molecules play the role of particles, a 1000-bit message can be transmitted by carefully releasing roughly 3000 molecules. Clearly, this requires very little energy and very little mass, which is ideal for nanoscale machines. In our normalized system where $d = \sigma^2 = 1$, it takes roughly 36000 seconds for the 3000 molecules to arrive, but a tiny value of $d$, which is appropriate for a nanoscale machine, would likely increase that rate significantly.

## V. Conclusion

This paper has explored the prospects for communication using particles that propagate across a medium using Brownian motion. Useful models and techniques have been derived which indicate that this is a feasible model for a communication system. However, much work remains to be done to create a practical system. Firstly, optimized input distributions need to be derived for these various methods. Furthermore, in the direction of Proposition 2, optimized tractable approximations need to be derived. Most importantly, practical methods of applying these results to communication in an actual nanoscale system need to be obtained, taking into account the complexity and energy constraints present in those systems.

## References

[1] S. P. Brown and R. A. Johnstone, "Cooperation in the dark: Signalling and collective action in quorum-sensing bacteria," *Proceedings of the Royal Society of London B*, vol. 268, pp. 961–965, 2001.

[2] R. Weiss and T. Knight, "Engineered communications for microbial robotics," in *Proc. 6th International Meeting on DNA Based Computers*, 2000.

[3] R. Weiss, S. Basu, S. Hooshangi, A. Kalmbach, D. Karig, R. Mehreja, and I. Netravali, "Genetic circuit building blocks for cellular computations, communications, and signal processing," *Natural Communication*, vol. 2, pp. 47–84, 2003.

[4] G. Chu, *System analysis of bacterial signalling*. M.A.Sc. thesis, University of Toronto, 2004.

[5] V. Anantharam and S. Verdú, "Bits through queues," *IEEE Trans. Inform. Theory*, vol. 42, pp. 4–18, Jan. 1996.

[6] R. Sundaresan and S. Verdú, "Capacity of queues via point-process channels," *IEEE Trans. Inform. Theory*, vol. 52, pp. 2697–2709, Jun. 2006.

[7] T. Berger, "Living information theory (Shannon lecture)," in *Proc. IEEE International Symposium on Information Theory*, Lausanne, Switzerland, 2002.

[8] L. G. Valiant, "The complexity of computing the permanent," *Theoretical Computer Science*, vol. 8, pp. 189–201, 1979.

[9] I. Karatzas and S. E. Shreve, *Brownian Motion and Stochastic Calculus (2nd edition)*. New York: Springer, 1991.

[10] C. Larédo and E. Klein, "A non-linear deconvolution problem coming from corn pollen dispersal estimation," in *Proc. Workshop on Dynamic Stochastic Modeling in Biology*, Copenhagen, Denmark, pp. 69–77, 2003.